\documentclass[preprint,aps,nofootinbib,showpacs,showkeys]{revtex4}

\usepackage{epsfig}
\usepackage{graphicx}
\usepackage{amssymb}
\usepackage{amsmath}
\usepackage{subfigure}
\usepackage{longtable}
\usepackage{epstopdf}

\begin{document}

\title{Low scale quantum gravity in gauge-Higgs unified models}

 \author{Jubin Park }
\email{honolov77@gmail.com}
%\vskip 0.5cm
\affiliation{Department of Physics, Chonnam National University,
300 Yongbong-dong, Buk-gu, Gwangju, 500-757, Republic of Korea}

%\date{\today}
\vspace{3cm}

\begin{abstract}
We consider the scale at which gravity becomes strong in linearized General Relativity coupled to the gauge-Higgs unified(GHU) model. We also discuss the unitarity of S-matrix in the same framework. The Kaluza-Klein(KK) gauge bosons, KK scalars and KK fermions in the GHU models can drastically change the strong gravity scale and the unitarity violation scale. In particular we consider two models $\mathrm{GHU_{\mathrm{SM}}}$ and $\mathrm{GHU_{\mathrm{MSSM}}}$ which have the zero modes corresponding to the particle content of the Standard Model and the Minimal Supersymmetric Standard Model, respectively. We find that the strong gravity scale could be lowered as much as $10^{13}(10^{14})$ GeV in the $\mathrm{GHU_{\mathrm{SM}}}$($\mathrm{GHU_{\mathrm{MSSM}}}$) for one extra dimension taking 1 TeV as the compactification scale. It is also shown that these scales are proportional to the inverse of the number of extra dimensions $d$. In the $d=10$ case, they could be lowered up to $10^{5}$ GeV for both models. We also find that the maximum compactification scales of extra dimensions quickly converge into one special scale $M_{O}$ near Planck scale or equivalently into one common radius $R_0$ irrespectively of $d$ as the number of zero modes increases. It may mean that all extra dimensions emerge with the same radius near Planck scale. In addition, it is shown that the supersymmetry can help to remove the discordance between the strong gravity scale and the unitarity violation scale.
\end{abstract}

% 11.10.Kk Field theories in dimensions other than four
% 11.15.-q Gauge field theories
% 11.25.Mj Compactification and four-dimensional models
% 04.60.Bc Phenomenology of quantum gravity
% 12.10.-g Unified field theories and models


\maketitle

%%%%%%%%%%%%%%%%%%%%%%%%%%%%%%%%%%%%%%%%%%%%%%%%%%%%%%%%%%%%%%%%%%%%%%%%
\section{Introduction}
The scale at which gravity becomes strong could be lowered as much as TeV scale which is much below the naively expected one (the reduced Planck mass) $\sim 10^{18}$ GeV. It is because a large non-minimal coupling of a single scalar field or Kaluza-Klein(KK) gravitons contribute to the renormalization group(RG) running of the reduced Planck mass~\cite{Atkins:2010eq,Atkins:2010re}. Moreover it is also well-known that the strong gravity scale could be different from the unitarity violation scale in linearized General Relativity coupled to matter~\cite{Han:2004wt}.
% Sometimes, it needs to be embedded into non-local models of quantum gravity.%

One important lesson from these recent studies is that the huge number of KK gravitons becomes a common source that lowers both of the scales, the strong gravity scale and the unitarity violation scale. For instance, in the large extra-dimensional model~\cite{ArkaniHamed:1998rs,Antoniadis:1998ig,Randall:1999ee}\footnote{The large extra dimension is introduced in order to solve the hierarchy problem by trading it for geometrical prescriptions such as the AdS geometry with a warping factor.} there exist $10^{32}$ KK gravitons. The low scale quantum gravity is expected and the unitarity violation occurs at a few hundred GeV. Therefore it is an appropriate time to question whether other sources like KK gravitons exist or not, and how they affect both of the scales. Keeping it in mind we focus on the gauge-Higgs unified(GHU) models~\cite{Manton:1979kb}\footnote{The electroweak scale is protected by a higher dimensional gauge symmetry.}, which naturally provide the KK gauge, KK scalar bosons (and KK fermions if bulk fermions are allowed). We show later that they can really change both of the scales depending on the number of extra dimensions $d$.

On the one hand, the $d$ is a crucial parameter in the GHU models. It constrains the structure of quartic terms of scalar potential.\footnote{The tree-level quadratic terms are also prohibited due to the shift symmetry. See ref.~\cite{Manton:1979kb} for one explicit example to generate quadratic terms in the monopole background.}
In the $d=1$, any quartic terms can not be generated at the tree-level in the scalar potential, while in the $d \geq 2$, tree-level quartic terms can be 
naturally generated from the commutators of zero 
modes in the field strength~\cite{Chang:2012iq}.
On the other hand, the $d$ significantly changes the total number of KK states. These increased KK states can lower the scale at which unitarity violates in the calculation of tree-level unitarity. More specifically, the partial-wave amplitude for a 2 $\rightarrow$ 2 elastic scattering~\cite{Han:2004wt} via s-channel graviton exchange is given by
\begin{equation}
a_2=-\frac{1}{40}G_N E_{CM}^2\, N~,\label{eq:Amplitude2}
\end{equation}
where $N \equiv \big(\frac{1}{3} N_S + N_F +4 N_V \big)$, and $N_S$, $N_F$, and $N_V$ are the number of real scalars, fermions and vector fields in the given model, respectively. Thus, the unitarity bound derived by $|\textrm{Re}~ a_J| \leq 1/2$ shows strong dependence on the total number of the KK states.
%%%%%%%%%%%%%%%%%%%%%%  Fscatter  %%%%%%%%%%%%%%%%%%%%%%%%%%%%%%%%
\begin{figure}%[h!t]
%\subfigure[]
\begin{center}
\includegraphics[width=0.45\textwidth]{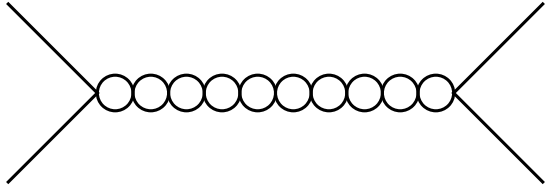}
%\subfigure[]
\caption{Scattering of elementary particles via s-channel graviton exchange.}\label{fig:figure1}
\end{center}
\end{figure}
%%%%%%%%%%%%%%%%%%%%%%%%%%%%%%%%%%%%%%%%%%%%%%%%%%%%%%%%%%%%%%%%%%

Generally there are two fundamental energy scales in the GHU models. As easily anticipated, one is the compactification scale$(\mathrm{M_C} \equiv 1/R)$ of extra dimensions, and the other one is the theory cutoff$(\Lambda_{\,\mathrm{CUTOFF}})$ from the effective field theory point of view.\footnote{Various experiments have been performed in order to search for deviations from Newton's law of gravitation, 
%\begin{equation}
$V(r)=-G_N \frac{m_1 m_2}{r} \big[ 1 + \alpha e^{-r/ \alpha} \big]~$.
%\end{equation}
See the ref.\,\cite{Adelberger:2009zz} for detailed explanation about experiments and current constraints for the compactification radius and scale.}\,\footnote{From now on, we assume that all extra dimensions have the same radius $\sim R$.}
In this letter we introduce one more scale parameter$(\Lambda_{\,\mathrm{UNIT}})$ reflecting the unitarity violation scale. Because of the hierarchy between the $\Lambda_{\,\mathrm{CUTOFF}}$ and the $\Lambda_{\,\mathrm{UNIT}}$ there may be some debate. We simply discuss it in the last part of Sec. II.

As two interesting benchmark models, we consider $\mathrm{GHU_{\mathrm{SM}}}$ and $\mathrm{GHU_{\mathrm{MSSM}}}$ which have the zero modes corresponding to the particle content of the Standard Model(SM) and the Minimal Supersymmetric Standard Model(MSSM), respectively. We find that the strong gravity scale could be lowered as much as a few hundred TeV. We also find that the supersymmetry not only make the maximum compactification scales of the extra dimensions converge into one special scale near Planck scale irrespectively of $d$, but also help to remove discordance between the strong gravity scale and the unitarity violation scale.

This paper is organized as follows. In Sec. II we briefly introduce the model, and show how to obtain these two scales $\Lambda_{\mathrm{CUTOFF}}$ and $\Lambda_{\mathrm{UNIT}}$. Next we consider aforementioned two models, $\mathrm{GHU_{\mathrm{SM}}}$ and $\mathrm{GHU_{\mathrm{MSSM}}}$ in order to show model-dependent results. Finally we analyze their numerical results, and discuss several scenarios depending on hierarchical patterns among three scales $\Lambda_{\mathrm{CUTOFF}}$, $\Lambda_{\mathrm{UNIT}}$, and $\mathrm{M_C}$. In Sec. III we summarize our paper.

%%%%%%%%%%%%%%%%%%%%%%%%%%%%%%%%%%%%% Table I %%%%%%%%%%%%%%%%%%%%%%%%%%%%%%%%%%%%%%%%%%%%%%%%%%%%%%
\begin{widetext}
%\squeezetable
\begin{center} % put inside center environment
\begin{table}[h!t]
\caption{Scattering amplitudes~\cite{Han:2004wt} for (complex) scalars, fermions, and vector bosons
via s-channel graviton exchange. They are written in terms of the Wigner $d^{\,(n)}_{n,m}$ functions~\cite{Beringer:1900zz}
in the massless limit. An overall factor $-2 \pi G_N E_{CM}^2$ has been
extracted from all amplitudes. The subscripts on the particles indicate
their helicities.}\label{tab:table1}
\begin{center}
\begin{tabular}{cccccc}
  \hline
  ~~~~~~~ $\longrightarrow$ ~~~~~~~&~~~~ $s's'$ ~~~~&~~~~ $\psi'_+\psi'_-$ ~~~~&~~~~ $\psi'_-\psi'_+$ ~~~~&~~~~ $V'_+V'_-$ ~~~~&~~~~ $V'_-V'_+$ \\
  \hline
  $s\bar{s}$~ & $\frac{2}{3}~d^{\,(2)}_{0,0}-\frac{2}{3}~(1+12\,\xi)^2~d^{\,(0)}_{0,0}$ & $\sqrt{2/3}~d^{\,(2)}_{0,1}$ & $\sqrt{2/3}~d^{\,(2)}_{0,-1}$
  & $2\sqrt{2/3}~d^{\,(2)}_{0,2}$ & $2\sqrt{2/3}~d^{\,(2)}_{0,-2}$ \\
%  \hline
  $f_+ \bar{f}_-$ & $\sqrt{2/3}~d^{\,(2)}_{1,0}$ & $d^{\,(2)}_{1,1}$ & $d^{\,(2)}_{1,-1}$ & $2d^{\,(2)}_{1,2}$ & $2d^{\,(2)}_{1,-2}$ \\
%  \hline
  $f_- \bar{f}_+$ & $\sqrt{2/3}~d^{\,(2)}_{-1,0}$ & $d^{\,(2)}_{-1,1}$ & $d^{\,(2)}_{-1,-1}$ & $2d^{\,(2)}_{-1,2}$ & $2d^{\,(2)}_{-1,-2}$ \\
%  \hline
  $V_+ V_-$ & $2\sqrt{2/3}~d^{\,(2)}_{2,0}$ & $2d^{\,(2)}_{2,1}$ & $2d^{\,(2)}_{2,-1}$ & $4d^{\,(2)}_{2,2}$ & $4d^{\,(2)}_{2,-2}$ \\
%  \hline
  $V_- V_+$ & $2\sqrt{2/3}~d^{\,(2)}_{-2,0}$ & $2d^{\,(2)}_{-2,1}$ & $2d^{\,(2)}_{-2,-1}$ & $4d^{\,(2)}_{-2,2}$ & $4d^{\,(2)}_{-2,-2}$ \\
  \hline
  \hline
\end{tabular}
\end{center}
\end{table}
\end{center}
\end{widetext}
%%%%%%%%%%%%%%%%%%%%%%%%%%%%%%%%%%%%%%%%%%%%%%%%%%%%%%%%%%%%%%%%%%%%%%%%%%%%%%%%%%%%%%%%%%%%%%%%%%%%%%%
\section{Model and fundamental energy scales}
The Lagrangian of linearized General Relativity coupled to particle content of the GHU model is given by
\begin{eqnarray}
S &=& \int d^{4}x\, \sqrt{-\,g} ~
\Big[ \frac{1}{16\pi G_N} (-2\lambda + R) \nonumber \\
&+& \Bigg( \frac{1}{2}g^{\mu\nu}\partial_{\mu}\phi^{\dagger}\partial_{\nu}\phi + \xi R \phi^2 + e\, \bar{\psi}i\gamma^{\mu}D_{\mu}\psi+\frac{1}{4}F_{\mu\nu}F^{\mu\nu}
\Bigg) \Big],
\end{eqnarray}
where $g$ is the determinant of the metric $g_{\mu \nu}$, $\lambda$ is the cosmological constant,
$R$ is the Ricci scalar, and $\xi$ is a free parameter. The scalar, fermion and vector fields in the Lagrangian stand for the typical fields of the GHU model. In particular, we focus on the non-minimal coupling case, $\xi=-1/12$, corresponding to the conformal limit of the theory~\cite{Callan:1970ze}. 
%It is consistent to the GHU model that has many KK scalar %fields because the $\xi \neq -1/12$ case can lead to a 5bound on the number of scalar fields~\cite{Atkins:%2010eq,Atkins:2010re}.

Now let us start by considering the s-channel scattering of matter particles via exchange of graviton.
These all amplitudes in the massless limit are represented in Table \ref{tab:table1}.
% for example : A(s sbar -> s s'bar)
%\begin{equation}
%\mathcal{A}(s \bar{s} \rightarrow s' \bar{s}')
%=-\frac{3}{4}\pi G_N E_{CM}^2 \Big(d^{\,(2)}_{0,0} -(1+12a)^2 d^{\,(0)}_{0,0} \Big)
%\end{equation}
The partial wave amplitude $a_J$ is extracted from $\mathcal{A}=16\pi \sum_J \, (2J + 1)\, a_J \,d^{\,(J)\,}_{\mu, \mu'}$.
In particular, each $J=0$ and $J=2$ partial wave amplitude can lead to the significant constraints to the $\Lambda_{\mathrm{UNIT}}$ scale and the matter content in the GHU models.
Note that the $J=0$ partial wave amplitude automatically vanishes due to $\xi=-1/12$ from $a_0 \sim (1+12\,\xi)^2$, while the $J=2$ partial waves do not change even if massive KK gravitons are involved~\cite{Atkins:2010re}.

As aforementioned, the large number of fields can induce a sizable running of the reduced Plank mass. More specifically, the RG equation for it is given by~\cite{Larsen:1995ax}
\begin{equation}
\overline{M}_P(\mu)^{\,2}=\overline{M}_P(0)^2 -\frac{1}{16\pi^2}\Big(\frac{1}{6}N_l+2\xi N_\xi \Big) \mu^2~,
\end{equation}
where $N_l \equiv \big( N_S+N_F-4N_{V} \big)$, $N_\xi$ is the number of real scalar fields non-minimally coupled to gravity, and $\mu$ is the renormalization scale.
In general, the strong gravity scale is evaluated when the fluctuations at length scale $\mu_{\star}$ is close to the reduced Planck scale $\overline{M}_P(\mu_{\star})$. We regard it as the cutoff $(\Lambda_{\mathrm{CUTOFF}})$ of the GHU models,
\begin{equation}
\mu_{\star}=\frac{\overline{M}_P(0)}{
\sqrt{1+\frac{1}{16\pi^2}\Big(\frac{1}{6}N_l+2\,\xi N_\xi \Big)}}\equiv \Lambda_{\mathrm{CUTOFF}} ~.\label{eq:LambdaCUTOFF}
\end{equation}

Before we discuss it in detail, it is worthwhile to mention an interesting relation which is induced by the boundary conditions on compact extra dimensions,
\footnote{Here we assume that $A_\mu\,(\mu=0, 1, 2, 3)$ and $A_i\,(i=5,6, \cdots)$ have the opposite boundary conditions of each other.}
\begin{equation}
\mathrm{GHU}~:~\mathcal{I} = N_S^{(0)} + N_V^{(0)}~,
\end{equation}
where superscripts $(0)$ denote zero modes for scalar and vector fields, and $\mathcal{I}$ is the number of generators of the original gauge group $G_M$.
For example, with $G_M=SU(3)$, if $G_M$ is broken into $SU(2)\times U(1)$, then we can have $``\,8 = 4 + (3 + 1)\,"$ relation, 
where the $4$ represents the (real) degrees of freedom of the Higgs doublet, and $8$, $3$, and $1$ correspond to the number of generators for each gauge generator of $SU(3)$, $SU(2)$ and $U(1)$, respectively.   
Therefore in general, two parameters $N_l^{(0)}$ and $N^{(0)}$ can be given in terms of $N_F^{(0)}$, $N_V^{(0)}$ and $\mathcal{I}$, 
\begin{equation}
N_l^{(0)} = \mathcal{I}+N_F^{(0)}-5N_{V}^{(0)} ~,~~ N^{(0)} = \frac{1}{3}\mathcal{I}+N_F^{(0)} +\frac{11}{3} N_V^{(0)} ~.
\end{equation}
Note that they can be used to remove degrees of freedom after fixing the $G_M$ and its branching rule to subgroups.

\begin{figure}%[h!t]
%\subfigure[aaa]
\begin{center}
\includegraphics[width=0.49\textwidth]{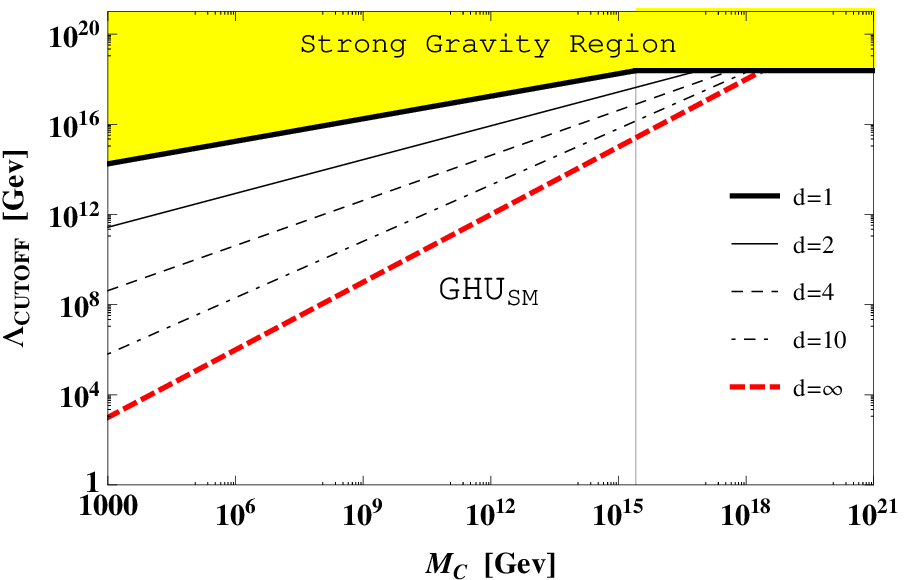}
\includegraphics[width=0.49\textwidth]{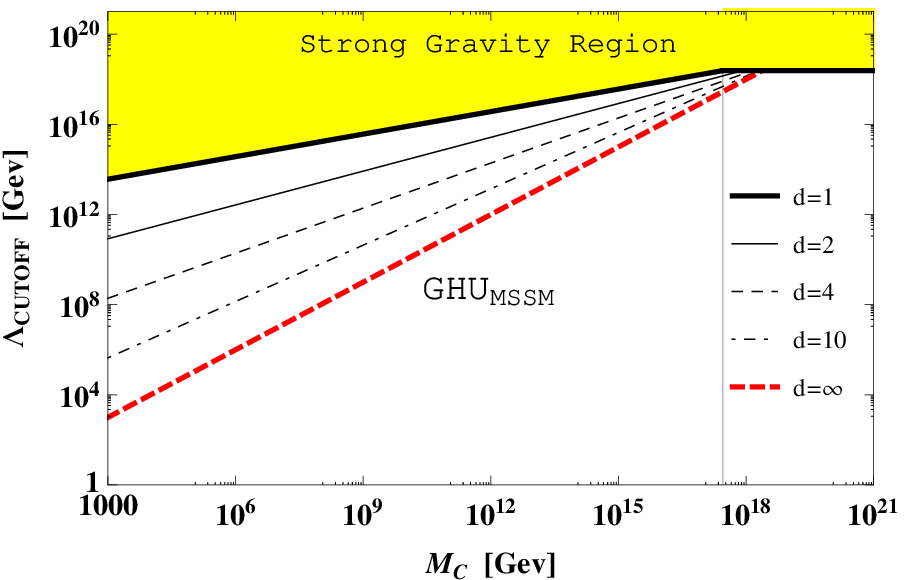}
%\subfigure[bbb]
\caption{$\Lambda_{\mathrm{CUTOFF}}$ as a function of $\mathrm{M_C}$. The left(right) panel is corresponding to the case of GHU model which has the degrees of freedom of the SM(MSSM). The $\Lambda_{\mathrm{CUTOFF}}$ is regarded as the scale at which gravity becomes strong. It also denotes that the perturbativity of the model breaks down. In the $d=1$ case the strong gravity region is painted yellow. The horizontal(vertical) line denotes the reduced Planck mass at $\mu=0$(the maximum $\mathrm{M_C}\, \equiv \mathrm{M_{max}}$).
Note that a red dashed line is corresponding to the $d=\infty$ case. It divides the $(\mathrm{M_C},\,\Lambda_{\mathrm{CUTOFF}})$ parameter space into two $\mathrm{M_C} < \Lambda_{\mathrm{CUTOFF}}$ (allowed region) and $\mathrm{M_C} > \Lambda_{\mathrm{CUTOFF}}$(forbidden region). 
}\label{fig:figure2}
\end{center}
\end{figure}

Again, let us turn back to the theory cutoff. After compactification, the GHU model becomes the 4-dimensional effective field theory with KK states of scalars and vector fields (and fermions if bulk fermions are allowed). Because they have mass spectra that have the same interval such as $1/R^2$, it is natural to assume that the total number of KK states of scalar$(S)$, vector$(V)$ and fermion$(F)$ fields is all the same,\,\footnote{For simplicity, we assume that our bulk space is flat. However in the warped (or curved) extra dimension, we should consider the red-shifted (or blue-shifted) energy spectrum. We do not consider it because it is beyond our present interest.}
\begin{equation}
N^{KK}_S= N^{KK}_V= N^{KK}_F \equiv \mathcal{J}_{KK}~.
\end{equation}
Note that the small differences among $S$, $V$ and $F$ modes due to boundary conditions are negligible because $N^{KK}_{X\, ,Y} \gg \Delta N^{(0)}_{XY}$, where $\Delta N^{(0)}_{XY}\equiv |N^{(0)}_X-N^{(0)}_Y|$ for $X,Y=\{S,V,F\}$. Thus, the cutoff scale in the GHU models is mainly dominated by the $\mathcal{J}_{KK}$ factor because $\mathcal{J}_{KK} \gg N_{\xi}, N_l^{(0)}$,
\begin{eqnarray}
\Lambda_{\mathrm{CUTOFF}} \sim  \frac{\bar{M}_P(0)}{
\sqrt{1+\mathcal{J}_{KK} N_l^{(0)} / (96 \pi^2)}}~,\label{eq:CUTOFF}
\end{eqnarray}
where the $N_l = \mathcal{J}_{KK}\, N_l^{(0)} = \mathcal{J}_{KK} \big( N_S^{(0)}+N_F^{(0)}-4N_V^{(0)} \big)$.
%For the reduced Planck mass M_P(0)=1/(8\pi G)=2.4353*10^{18} GeV%
In addition, the number of KK states with $d$ extra dimensions is easily calculated by
\begin{equation}
\mathcal{J}_{KK} \sim \Big( \frac{\Lambda_{\mathrm{CUTOFF}}}{1/R}\Big)^d=\Big(\frac{\Lambda_{\mathrm{CUTOFF}}}{\mathrm{M_C}}\Big)^d~.\label{def:KK}
\end{equation}
The $\Lambda_{\mathrm{CUTOFF}}$ as a function of $\mathrm{M_C}$ is obtained with the above two relations (neglecting a constant $1$ in a denominator of Eq.\,(\ref{eq:CUTOFF})\,)
\begin{equation}
\Lambda_{\mathrm{CUTOFF}}=\Bigg[\frac{\mathrm{M_C}^d\, \bar{M}^2_P(0)}{N_l^{(0)}/ 96\,\pi^2}\Bigg]^{1/(2+d)}~.
\label{eq:CUTOFF2}
\end{equation}
%%%%%%%%%%%%%%%%%%%%%%%%%%%% Table %%%%%%%%%%%%%%%%%%%%%%%%%%%%%%
\begin{widetext}
%\squeezetable
\begin{center} % put inside center environment
\begin{table}%[h!t]
\begin{center} % put inside center environment
\caption{The cutoff scale $(\Lambda_{\mathrm{CUTOFF}})$ of the GHU model that has the zero modes corresponding to the particle content of the SM(or MSSM) at $\mathrm{M_C}=1$ TeV. The $d$ and $\mathrm{M_{max}}$ denote the number of extra dimensions and the maximum $\mathrm{M_C}$, respectively. Note that the $\mathrm{M_{max}}$ may be regarded as the upper bound of $\mathrm{M_C}$ (see the main body). As the $d$ increases, the $\Lambda_{\mathrm{CUTOFF}}\,(\mathrm{M_C} = 1 \mathrm{TeV})$ drastically decreases. On the contrary, the $\mathrm{M_{max}}$ slowly increases until the $\Lambda_{\mathrm{CUTOFF}}$ is equal to the reduced Planck mass at $\mu=0$.} 
\label{tab:table2}
\begin{tabular}{|c|c|c|c|c|}
 \hline
 & \multicolumn{2}{c|} {$\mathrm{GHU_{\mathrm{SM}}}$} &
    \multicolumn{2}{c|} {$\mathrm{GHU_{\mathrm{MSSM}}}$} \\
  \hline
  $d$ & $\Lambda_{\mathrm{CUTOFF}}(\mathrm{M_C} = 1$ TeV)\,[Gev] & $\mathrm{M_{max}}$\,[Gev] & $\Lambda_{\mathrm{CUTOFF}}(\mathrm{M_C} = 1$ TeV)\,[GeV] & $\mathrm{M_{max}}$\,[GeV] \\
  \hline
  ~~~1~~~ & ~$1.78\times 10^{14}$~ & ~~~$2.57\times 10^{15}$~~~ & ~~~$3.70\times 10^{13}$~~~&~~~ $2.85 \times 10^{17}$~~~\\
  2 & $2.74\times 10^{11}$ & $7.91 \times 10^{16}$ &$8.44 \times 10^{10}$& $8.34 \times 10^{17}$\\
  4 & $4.21\times 10^{8}$ & $4.39\times 10^{17}$ & $1.92\times 10^{8}$& $1.42\times 10^{18}$ \\
  10 & $6.49\times 10^{5}$ & $1.23\times 10^{18}$ & $4.39 \times 10^{5}$& $1.97\times 10^{18}$ \\
  $\infty$ & $10^{3}$ & $2.44\times 10^{18}$ & $10^{3}$ & $2.44\times 10^{18}$ \\
  \hline
\end{tabular}
\end{center}
\end{table}
\end{center}
\end{widetext}
%%%%%%%%%%%%%%%%%%%%%%%%%%%%%%%%%%%%%%%%%%%%%%%%%%%%%%%%%%%%%%%%%%
Numerically, $N_l^{(0)}=1$
for the SM which has $N_S=4$, $N_F=45$ and $N_V=12$. For the MSSM which has two Higgs doublets, $N_S=98$, $N_F=61$ and $N_V=12$, the $N_l^{(0)}=111$. The $\Lambda_{\mathrm{CUTOFF}}$ for $\mathrm{GHU_{\mathrm{SM}}}$ and $\mathrm{GHU_{\mathrm{MSSM}}}$ at the $\mathrm{M_C}=1$ TeV is calculated by
\begin{eqnarray}
&&\Lambda_{\mathrm{CUTOFF}}^{\mathrm{SM}} \sim \Big(5.62\times 10^{(39+3d)}\Big)^{1/(2+d)} \nonumber ~,\\
&&\Lambda_{\mathrm{CUTOFF}}^{\mathrm{MSSM}} \sim \Big(5.06\times 10^{(37+3d)}\Big)^{1/(2+d)}~.
\end{eqnarray}
We present numerical results of the $\Lambda_{\mathrm{CUTOFF}}$ for both models in Table \ref{tab:table2}. In Table \ref{tab:table2}, the first column $d$ denotes the number of extra dimensions, and the second and the fourth columns show the cutoff scales at $\mathrm{M_C}=1$ TeV. Interestingly, they show that the strong gravity scale could be much lower than the reduced Planck mass $\sim 10^{18}$ GeV, and it could appear at $10^{5} \sim 10^{13}$ or $10^{14}$ GeV depending on $d=10 \sim 1$. Additionally, the third and the fifth columns denote the maximum $\mathrm{M_C}$ $(\mathrm{M_{max}})$ when the cutoff scale (as a function of $\mathrm{M_C}$) is equal to the reduced Planck mass $\bar{M}_P(\mu=0)$ by varying the $\mathrm{M_C}$ from $10^3$ GeV to $10^{21}$ GeV (see maximum points around vertical lines in both panels in Fig.\,\ref{fig:figure2}). 
%%%%%%%%%%%%%%%%%%%% Figure 2 %%%%%%%%%%%%%%%%%%%%%%%%%%%%%%%%%%%%
%\begin{widetext}
%\begin{center}
%\end{center}
%\end{widetext}

We also plot the $\Lambda_{\rm{CUTOFF}}$ as a function of $\mathrm{M_C}$ with a fixed number of $d$ in Fig.\,\ref{fig:figure2}. The Left(right) panel is corresponding to the case of $\mathrm{GHU}_{\mathrm{SM}}$  $(\mathrm{GHU}_\mathrm{MSSM})$. In each panel, we choose the $d=1$ case as a reference case. Its strong gravity region is painted yellow. Additionally the horizontal and vertical lines in both panels are used to denote the reduced Planck mass at $\mu=0$ and the $\mathrm{M_{max}}$ when $d=1$, respectively. Note that when $\mathrm{M_C} > \mathrm{M_{max}}$, it seems that the $\Lambda_{\mathrm{CUTOFF}}$ can be larger than the horizontal line $\bar{M}_P(\mu=0)$. However, it is not consistent because 
%$\bar{M}_P(\mu \neq 0) < \bar{M}_P(\mu=0)$ unless $\xi < 0$, see %Eq.\,(\ref{eq:LambdaCUTOFF}) or Eq.\,(\ref{eq:CUTOFF}). In fact, 
the enhancement to the $\bar{M}_P(\mu=0)$ is not allowed due to the constant $1$ in a denominator of Eq.\,(\ref{eq:CUTOFF}). 
%$\big($or even the $\mathcal{J}_{KK} N_l^{(0)} / (96 \pi^2) < 1$ %when $\mathrm{M_C} > \mathrm{M_{max}}$ $\big)$.
%\footnote{Note that Eq.\,(\ref{eq:CUTOFF}) does not correctly work %with the small number of KK states. %because the Eq.\,%(\ref{eq:CUTOFF2}) shows the enhancement %to $%\Lambda_{\mathrm{CUTOFF}}$ under it at $\mathrm{M_C} > %\mathrm{M_{max}}$%$.} 
Finally, the red dashed line in Fig.\,\ref{fig:figure2} is corresponding to the $d=\infty$ case. It divides the $(\mathrm{M_C},\,\Lambda_{\mathrm{CUTOFF}})$ parameter space into the $\mathrm{M_C} < \Lambda_{\mathrm{CUTOFF}}$ region(allowed region) and the $\mathrm{M_C} > \Lambda_{\mathrm{CUTOFF}}$ region(forbidden region). 

We find two interesting facts from the above numerical analysis. Firstly, there exists a tension between $d$ and $\Lambda_{\mathrm{CUTOFF}}$, that is to say, when the $d$ increases, the $\Lambda_{\mathrm{CUTOFF}}$ drastically decreases, and vice-verse. Interestingly, the $d=10$ case shows that the $\Lambda_{\mathrm{CUTOFF}}$ could be lowered to a few hundred TeV at $M_C = 1$ TeV.\footnote{On the other hand, it implies that the $d > 10$ case could be excluded from negative experimental data about the low scale quantum gravity below a few hundred TeV in gravitational and collider experiments.}  Secondly, as the number of zero modes increases (for example, $N^{(0)}_{SM}\,$ of the SM $\rightarrow$ $N^{(0)}_{MSSM}\,$ of the MSSM), it seems that the maximum compactification scales $(\mathrm{M_{max}})$ quickly converge into one special scale (see around the vertical line in the right panel in Fig.\,\ref{fig:figure2}).
It is very intriguing that any $\mathrm{GHU}_{\mathrm{MSSM}}$  with an arbitrary $d$ finally has one common $\mathrm{M_{max}}$ near the reduced Planck mass. Actually, we find that all lines meet at one scale near Planck scale (from now on, let us call it $``M_{O}"$ or equivalently $``R_0"$ as one common compactification radius).
It may mean that all extra dimensions emerge with the same radius near Planck scale, while the extra dimensions which have $\mathrm{M_C} > M_O$ or $R < R_0$ rapidly dissolve in the strong gravity region. In this sense, we may say that all compactification radii of extra dimensions are unified at $M_O$. Note that this situation is analogous to the unification of gauge couplings in the MSSM. Therefore, the supersymmetry could not only unify the gauge couplings but also unify all radii of extra dimensions into the $R_0$ near Planck scale.

\begin{figure}%[h!t]
%\subfigure[aaa]
\begin{center}
\includegraphics[width=0.49\textwidth]{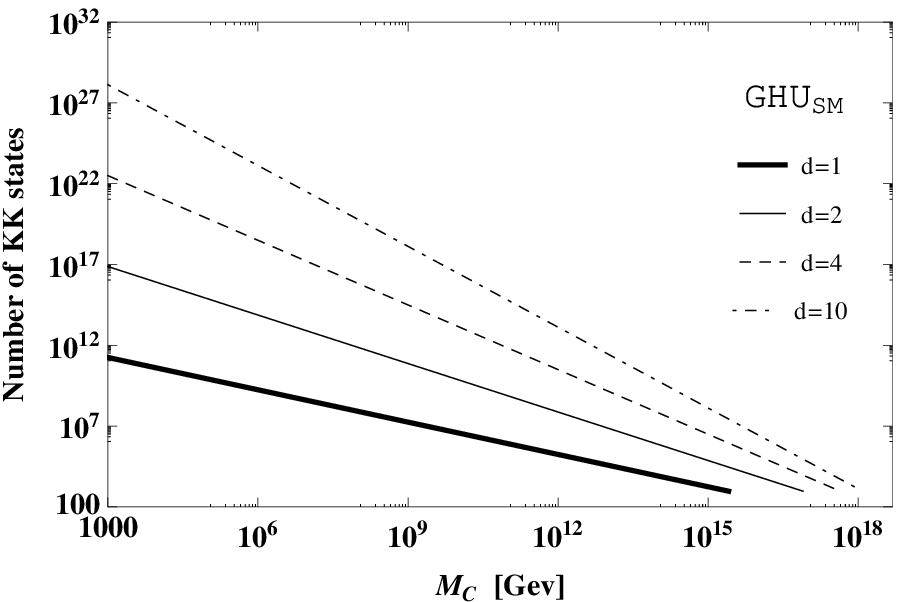}
\includegraphics[width=0.49\textwidth]{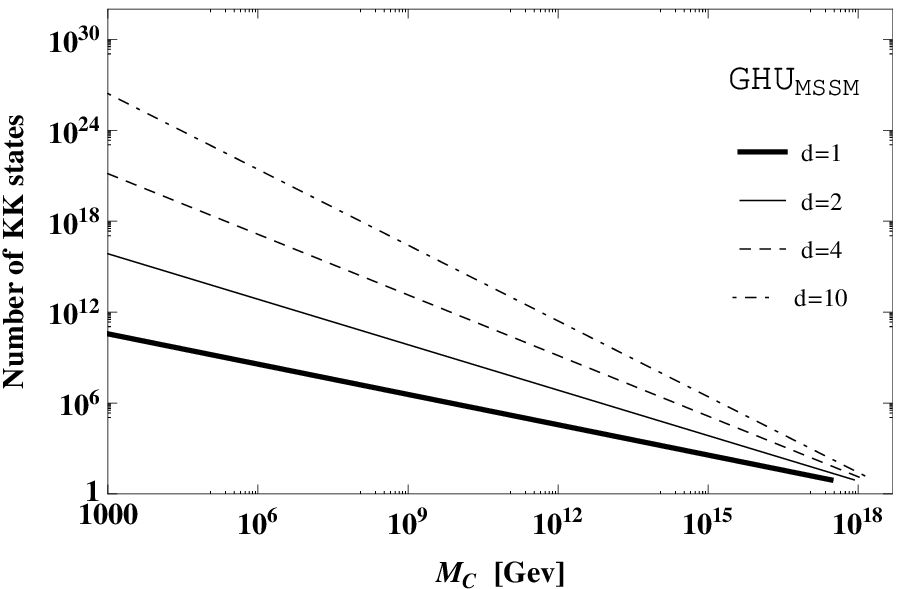}
%\subfigure[bbb]
\caption{The number of KK states $\mathcal{J}_{KK}$ as a function of compactification scale $\mathrm{M_C}$. The number $d$ denotes the number of extra dimensions. The left(right) panel corresponds to the case of the $\mathrm{GHU}_{\mathrm{SM}}\,(\mathrm{GHU}_{\mathrm{MSSM}})$. These $\mathcal{J}_{KK}$ numbers drastically decrease as the $\mathrm{M_C}$ scale increases because $\mathcal{J}_{KK} \sim \big(\Lambda_{\mathrm{CUTOFF}} / (1/R) \big)^d= \big(\Lambda_{\mathrm{CUTOFF}}/ \mathrm{M_C}\big)^d$.
}\label{fig:figure3}
\end{center}
\end{figure}

Now let us turn out our attention into the $J=2$ partial wave amplitude. Because it has additional overall factors due to the degrees of freedom of KK states $\big($see Eq.\,(\ref{eq:Amplitude2}) for the original amplitude $\big)$, it has this general form of
\begin{equation}
a_2=-\frac{1}{40}G_N E_{CM}^2 \mathcal{J}_{KK}^{G} \mathcal{J}_{KK} N^{(0)}~,
\end{equation}
where $\mathcal{J}_{KK}^{G}$ is the total number of KK gravitons, and $N^{(0)}=\big(\frac{1}{3} N_S^{(0)} + N_F^{(0)} +4 N_V^{(0)} \big)$.
For one instructive example, let us consider the large extra dimensions scenario~%\footnote{$M_{Pl}^2=M_{*}^{2+d}\,V^{d}$} 
where the gravitons propagate in the bulk, while all matter and gauge fields are confined to the 3-dimensional membrane. In this case we have the $\mathcal{J}_{KK}^{G} = 10^{\,32}$ and the $\mathcal{J}_{KK}=1$~\cite{Atkins:2010eq,Atkins:2010re}.
By applying the unitarity condition $|a_2| \leq 1/2$, the energy scale at which
tree-level unitarity violates is given by
\begin{equation}
E_{CM}^2 = \frac{20}{G_N\,N^{(0)}} \frac{1}{\mathcal{J}_{KK}^{G}\,\mathcal{J}_{KK}} = \frac{E_{CM}^{(0)~2}}{\mathcal{J}_{KK}^{G}\,\mathcal{J}_{KK}}~~ \label{eq:Lambda_UNIT}
\end{equation}
where $E_{CM}^{(0)~2} \equiv 20 (G_N\,N^{(0)})^{-1}$.
Numerically, $E_{CM}^{(0)} \approx 6 \times 10^{18}$ GeV for the SM, and $E_{CM}^{(0)} \approx 4 \times 10^{18}$ GeV for the MSSM.
The unitarity violation in the large extra dimensions scenario thus occurs at the $E_{CM}$,
\begin{eqnarray}
&&\Lambda_{\mathrm{UNIT}}^{\mathrm{\,SM}}\equiv E_{CM} \sim \sqrt{\frac{(6 \times 10^{18})^2}{ 10^{32} \times 1}} = 600~\mathrm{GeV}~,\nonumber \\
&&\Lambda_{\mathrm{UNIT}}^{\mathrm{MSSM}} \sim  400~\mathrm{GeV}~,
\end{eqnarray}
Note that they are approximate estimates due to the massless limit of KK gravitons (See Ref. \cite{Atkins:2010re} for more exact numbers). Similarly, many KK states of scalar, vector and fermion fields in the context of GHU models behave like KK gravitons when considering the theory cutoff and the unitarity. As aforementioned, we introduce another parameter $\Lambda_{\,\mathrm{UNIT}}$ reflecting the unitarity violation scale.
Because the $N_{KK}$ is in inverse proportion to $\mathrm{M_C}$ (see Eq.\;(\ref{eq:Lambda_UNIT})), the $\Lambda_{\mathrm{UNIT}}\equiv E_{CM}$ is proportional to $\mathrm{M_C}$. Namely, if the $\mathrm{M_C}$ increases, the number of KK states decreases and it can raise the scale of unitarity violation, while if the $\mathrm{M_C}$ decreases, then the $N_{KK}$ increases and the $\Lambda_{\mathrm{UNIT}}$ decreases.
Numerically, if we take $\mathrm{M_C} = 1$ TeV with $d=1$, then $\Lambda_{\mathrm{CUTOFF}}^\mathrm{GHU_{\mathrm{SM}}}=1.78 \times 10^{14}$ GeV (see Table \;\ref{tab:table2}) and the number of KK states is
\begin{equation}
\mathcal{J}_{KK} \sim \Bigg(\frac{1.78\times 10^{14}}{10^{3}}\Bigg)^{1}=1.78 \times 10^{11}~.\label{eq:KK}
\end{equation}
%where we ignore some constant factor in energy formula of KK states except $1/R$ because they are negligible compared with $\mathcal{J}_{KK}$ itself. 
We plot the $\mathcal{J}_{KK}$ as a function of $\mathrm{M_C}$ in Fig.\,\ref{fig:figure3} for both $\mathrm{GHU}_{\mathrm{SM}}$(left panel) and $\mathrm{GHU}_{\mathrm{MSSM}}$(right panel). They show that the $\mathcal{J}_{KK}$ drastically decreases as the $\mathrm{M_C}$ increases. 

With this $\mathcal{J}_{KK}$, the $\Lambda_{\mathrm{UNIT}}^{\mathrm{GHU_{SM}}}$ is easily calculated by
\begin{equation}
\Lambda_{\mathrm{UNIT}}^{\mathrm{GHU_{SM}}} \sim \sqrt{\frac{(6 \times 10^{18})^2}{1.78 \times 10^{11}}} \sim 1.42\times 10^{13} ~\mathrm{GeV}.\label{eq:E_CM}
\end{equation}
It is interesting that the theory cutoff and the unitarity violation scale do not coincide in the $\mathrm{GHU_{SM}}$ ($\Lambda_{\mathrm{CUTOFF}}^{\mathrm{GHU_{SM}}} \neq \Lambda_{\mathrm{UNIT}}^{\mathrm{GHU_{SM}}}$). In the same way, we calculate the $\mathcal{J}_{KK}$ and the $\Lambda_{\mathrm{UNIT}}$ by varying $d$ from 0 to $\infty$. These numerical results are presented in Table \ref{tab:table3}. As the $d$ increases, the $\mathcal{J}_{KK}$ rapidly increase and the $\Lambda_{\mathrm{UNIT}}$
drastically decreases. In particular, the $d=10$ case shows that the unitarity violation scale could be lowered as much as $\sim 10$ TeV similarly to the previous case of the theory cutoff. It is also found that $\Lambda_{\mathrm{CUTOFF}}^{\mathrm{GHU_{SM}}} > \Lambda_{\mathrm{UNIT}}^{\mathrm{GHU_{SM}}}$ in the $\mathrm{GHU_{SM}}$, while $\Lambda_{\mathrm{CUTOFF}}^{\mathrm{GHU_{MSSM}}} \approx \Lambda_{\mathrm{UNIT}}^{\mathrm{GHU_{MSSM}}}$ in the $\mathrm{GHU_{MSSM}}$. It is thus expected that there is different physics at around $\Lambda_{\mathrm{UNIT}}$ in each model. In the following subsections, we discuss several scenarios depending on the hierarchical patterns among $\Lambda_{\mathrm{UNIT}}$, $\Lambda_{\mathrm{CUTOFF}}$, and $\mathrm{M_C}$. 

%%%%%%%%%%%%%%%%%%%%% Figure 3 %%%%%%%%%%%%%%%%%%%%%%%%%%%%%%%%%%%%
%\begin{widetext}
%\begin{center}
%\begin{center}
%\end{center}
%\end{widetext}
%%%%%%%%%%%%%%%%%%%%%%%%%%%%%%%%%%%%%%%%%%%%%%%%%%%%%%%%%%%%%%%%%%%%%%%%%%%%%%%

%%%%%%%%%%%%%%%%%%%%%%%%%%%%%%%%%%%%%%%%%%%%%%%%%%%%%%%%%%%%%%%%%%%%%%%%%%%%%%%
\begin{widetext}
%\squeezetable
\begin{center} % put inside center environment
\begin{table}%[h!t]
\begin{center} % put inside center environment
\caption{The $d$\,, $\mathcal{J}_{KK}$, $\Lambda_{\mathrm{UNIT}}$ and the radius of extra dimensions for both $\mathrm{GHU_{\mathrm{SM}}}$ and $\mathrm{GHU_{\mathrm{MSSM}}}$ are presented in sequence. They are evaluated by taking $\mathrm{M_C}=1$ TeV and the $\Lambda_{\mathrm{CUTOFF}}$ values in Table \ref{tab:table2}. 
% After fixing \Lambda_CUTOFF = 10^18, 
% $d \geq 4$ extra dimensions are not consistent from experimental % constraints, $R \leq 44 ~ \mu m$~\cite{Adelberger:2009zz}, 
% equivalently $1/R ~\geq~ 4.5 \times 10^{-3}$ eV. 
Interestingly, numerical results show that $\Lambda_{\mathrm{CUTOFF}} > \Lambda_{\mathrm{UNIT}}$ in the $\mathrm{GHU_{\mathrm{SM}}}$, while $\Lambda_{\mathrm{CUTOFF}} \approx \Lambda_{\mathrm{UNIT}}$ in the $\mathrm{GHU_{\mathrm{MSSM}}}$. In addition the $d=10$ case shows very low unitarity violation scales $\sim 10^{4}\,(\mathrm{GHU_{\mathrm{SM}}})$, $10^5\,(\mathrm{GHU_{\mathrm{MSSM}}})$ GeV. $\big[$For one reference, we present the experimental constraint of gravitation, $R \leq 44 ~ \mu m$~\cite{Adelberger:2009zz} or equivalently $1/R ~\geq~ 4.5 \times 10^{-3}$ eV, and the collider constraint $\mathrm{M_C} > 1.59$ TeV with CL=95\% from $p\,\bar{p} \rightarrow $ dijet, angular distrib.~\cite{Beringer:1900zz} $\big]$}\label{tab:table3}
\begin{tabular}{|c|c|c|c|c|c|c|}
  \hline
  & \multicolumn{3}{c|} {$\mathrm{GHU_{\mathrm{SM}}}$} &
    \multicolumn{3}{c|} {$\mathrm{GHU_{\mathrm{MSSM}}}$} \\
  \hline
  $d$ & $\mathcal{J}_{KK}$ & $\Lambda_{\mathrm{UNIT}}$ [GeV] &  Radius[m] & $\mathcal{J}_{KK}$ & $\Lambda_{\mathrm{UNIT}}$ [eV] &  Radius[m] \\
  \hline
  ~~1~~ &~~ $1.78\times 10^{11}$ ~~&~~ $1.42\times 10^{13}$ ~~&~~ $1.38\times 10^{-25}$ ~~&~~ $3.70\times 10^{10}$ ~~&~~ $2.08 \times 10^{13}$ ~~&~~ $9.49\times 10^{-26}$ \\
  2 & $7.51\times 10^{16}$ & $2.19\times 10^{10}$ & $9.01\times 10^{-23}$ & $7.12 \times 10^{15}$ & $4.74 \times 10^{10}$ & $4.16\times 10^{-23}$\\
  4 & $3.14\times 10^{22}$ & $3.39\times 10^{7}$ & $5.83\times 10^{-20}$ & $1.36\times 10^{21}$ & $1.09\times 10^{8}$ & $1.82\times 10^{-20}$ \\
  10 & $1.33\times 10^{28}$ & $5.21\times 10^{4}$ & $3.79\times 10^{-17}$ & $2.66 \times 10^{26}$ & $2.45\times 10^{5}$ & $8.04\times 10^{-18}$\\
  $\infty$ & $1$ & $10^{3}$ & $1.97 \times 10^{-15}$ & 1 & $10^3$ & $1.97 \times 10^{-15}$\\
  \hline
\end{tabular}
\end{center}
\end{table}
\end{center}
\end{widetext}
%%%%%%%%%%%%%%%%%%%%%%%%%%%%%%%%%%%%%%%%%%%%%%%%%%%%%%%%%%%%%%%%%%%

%%%%%%%%%%%%%%%%%%%%%%%%%%%%%%%%%%%%%%%%%%%%%%%%%%%%%%%%%%%%%%%%%%%%

\subsection{$\Lambda_{\mathrm{UNIT}} > \Lambda_{\mathrm{CUTOFF}} $}
The theory enters into the strong interaction region above $\Lambda_{\mathrm{UNIT}}$ scale because the perturbativity of the model breaks down. The (perturbative) effective field theory remains valid below this scale. However it is not consistent because the $\Lambda_{\mathrm{CUTOFF}}$ is already smaller than the $\Lambda_{\mathrm{UNIT}}$.

\subsection{$\Lambda_{\mathrm{UNIT}} < \Lambda_{\mathrm{CUTOFF}}$}
In this case, there exists an intermediate energy gap between the strong gravity scale and the unitarity violation scale $\big($see Fig.\;\ref{fig:figure4}\;(a)$\big)$. In order to make the theory consistent we should assume some mechanism or new physics that can restore the unitarity. Actually, this scenario happens in the $\mathrm{GHU}_{\mathrm{SM}}$. Here the $\Lambda_{\mathrm{CUTOFF}}$ is about ten times larger than $\Lambda_{\mathrm{UNIT}}$ (see Table \ref{tab:table2} and Table \ref{tab:table3}). As one candidate of new physics, the stringy effects may help to remedy the unitarity violation. If it really happens, they may leave some new physics signals at that scale.

\subsection{$\Lambda_{\mathrm{UNIT}} \approx \Lambda_{\mathrm{CUTOFF}}$}
In the $\mathrm{GHU}_{\mathrm{MSSM}}$ case, this scenario is realized $\big($see Fig.~\ref{fig:figure4}~(b)$\big)$. There is no  unnatural discordance between the $\Lambda_{\mathrm{UNIT}}$ and the $\Lambda_{\mathrm{CUTOFF}}$ scales. Any new physics is not needed in order to remedy the unitarity violation. It is worthwhile to recall that the zero modes increased by supersymmetry can reduce the gap between the $\Lambda_{\mathrm{UNIT}}$ and the $\Lambda_{\mathrm{CUTOFF}}$ scales
due to the reduced $E_{CM}^{(0)}$ and increased KK numbers in the Eq.\,(\ref{eq:Lambda_UNIT}). 

\subsection{$\Lambda_{\mathrm{UNIT}} \approx \Lambda_{\mathrm{CUTOFF}} \approx \mathrm{M_C}$}
In this scenario, there exists only one new physics scale. It is thus impossible to have any KK states except zero modes because there is no room for them. Whole spectrum consists of all zero modes. Consequently the effective GHU models may not be distinguishable from the SM if there are no additional zero modes.

%%%%%%%%%%%%%%%%%%%%%% Figure 4 %%%%%%%%%%%%%%%%%%%%%%%%%%%%%%%%%%%
\begin{widetext}
%\squeezetable
\begin{center}
\begin{figure}%[h!t]
%\subfigure[aaa]
\begin{center}
\includegraphics{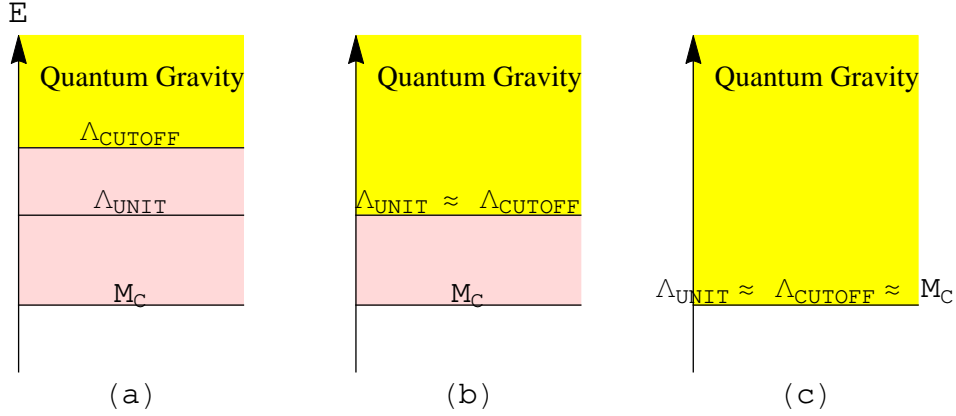}
%\subfigure[bbb]
\caption{Schematic illustration of energy scales $\big(\Lambda_{\,\mathrm{CUTOFF}},\, \Lambda_{\,\mathrm{UNIT}},\, \mathrm{\mathrm{M_C}}\big)$ in linearized General Relativity coupled to the gauge-Higgs unified(GHU) model.
%where $\Lambda_{\,\mathrm{CUTOFF}}$, $\Lambda_{\,\mathrm{UNIT}}$, %and $\mathrm{\mathrm{M_C}}$ denote theory cutoff, the scale of unitarity %violation and compactification scale, respectively. 
The yellow regions show the quantum gravity region. The red regions correspond to the weak gravity region in which KK states can live. The left panel (a) shows that there exists the discordance between $\Lambda_{\,\mathrm{CUTOFF}}$ and $\Lambda_{\,\mathrm{UNIT}}$.
%Here some new physics such as string theory should be introduced %in order to recover the unitarity and draw the scale to %$\Lambda_{\,\mathrm{CUTOFF}}$.
The central panel (b) shows that the $\Lambda_{\,\mathrm{UNIT}}$ has the same order of magnitude as the $\Lambda_{\,\mathrm{CUTOFF}}$ scale. The right panel (c) shows that three scales coincide.
It is thus expected that there is different physics at around the $\Lambda_{\mathrm{UNIT}}$ scale in each case.}\label{fig:figure4}
\end{center}
\end{figure}
\end{center}
\end{widetext}
%%%%%%%%%%%%%%%%%%%%%%%%%%%%%%%%%%%%%%%%%%%%%%%%%%%%%%%%%%%%%%%%
\section{Conclusion}
In summary, we have studied the strong gravity scale and the unitarity violation scale in linearized General Relativity coupled to particle content of the GHU model. The KK gauge bosons, KK scalars and KK fermions in the GHU models drastically change both of the scales. In particular we have considered the two interesting benchmark models, $\mathrm{GHU_{\mathrm{SM}}}$ and $\mathrm{GHU_{\mathrm{MSSM}}}$ in order to show model-dependent difference. We have found that the strong gravity scale could be lowered as much as $10^{13}(10^{14})$ GeV in the $\mathrm{GHU_{\mathrm{SM}}}$($\mathrm{GHU_{\mathrm{MSSM}}}$) by taking $\mathrm{M_C}=1$ TeV and $d=1$. It is also shown that these scales are proportional to the inverse of $d$. In the $d=10$ case, they could be lowered up to $10^{5}$ GeV for both of the models.

We have also found that the maximum compactification scales $(\mathrm{M_{max}})$ of extra dimensions quickly converge into one special scale $``M_{O}"$ near Planck scale or equivalently into one common radius $``R_0"$ irrespectively of $d$, when the number of zero modes increases (for example, $N^{(0)}_{\mathrm{SM}} \rightarrow N^{(0)}_{\mathrm{MSSM}})$. It may mean that there is the unification of compactification radii near Planck scale analogously to the unification of gauge couplings in the MSSM. Moreover, it is also interesting that the supersymmetry helps to remove the discordance between the $\Lambda_{\mathrm{UNIT}}$ and the $\Lambda_{\mathrm{CUTOFF}}$ scales. Consequently, it may reveal that the supersymmetry can play another important role in extra dimensions. 

Finally, our method can be easily applied to the other extra dimensional models that have these KK states.

\begin{acknowledgements}
J. P was supported by the National Research Foundation of Korea (NRF) grant (No. 2013R1A2A2A01015406). J. P thanks J.S. Lee for his valuable comments.
\end{acknowledgements}

\end{document}